\documentclass[pra,twocolumn,aps,superscriptaddress]{revtex4}

\usepackage{times}
\usepackage{graphicx}
\usepackage{epsfig}
\usepackage{amsfonts}
\usepackage{amsmath}
\usepackage{amssymb}
\usepackage{color}
\usepackage{multirow}
\usepackage[colorlinks=true,citecolor=blue,urlcolor=blue]{hyperref}
\usepackage{float}

\begin{document}


\title{Experimental demonstration of the connection between quantum contextuality and graph theory}



\date{\today}


\author{Gustavo Ca\~{n}as}
\affiliation{Departamento de F\'{\i}sica, Universidad de Concepci\'on, 160-C Concepci\'on, Chile}
\affiliation{Center for Optics and Photonics, Universidad de Concepci\'on, Concepci\'on, Chile}
\affiliation{MSI-Nucleus for Advanced Optics, Universidad de Concepci\'on, Concepci\'on, Chile}
\email{gcanasc@udec.cl}

\author{Evelyn Acu\~{n}a}
\affiliation{Departamento de F\'{\i}sica, Universidad de Concepci\'on, 160-C Concepci\'on, Chile}
\affiliation{Center for Optics and Photonics, Universidad de Concepci\'on, Concepci\'on, Chile}
\affiliation{MSI-Nucleus for Advanced Optics, Universidad de Concepci\'on, Concepci\'on, Chile}

\author{Jaime Cari\~{n}e}
\affiliation{Center for Optics and Photonics, Universidad de Concepci\'on, Concepci\'on, Chile}
\affiliation{MSI-Nucleus for Advanced Optics, Universidad de Concepci\'on, Concepci\'on, Chile}
\affiliation{Departamento de Ingenier\'{\i}a El\'ectrica, Universidad de Concepci\'on, 160-C Concepci\'on, Chile}

\author{Johanna F. Barra}
\affiliation{Departamento de F\'{\i}sica, Universidad de Concepci\'on, 160-C Concepci\'on, Chile}
\affiliation{Center for Optics and Photonics, Universidad de Concepci\'on, Concepci\'on, Chile}
\affiliation{MSI-Nucleus for Advanced Optics, Universidad de Concepci\'on, Concepci\'on, Chile}

\author{Esteban S. G\'omez}
\affiliation{Departamento de F\'{\i}sica, Universidad de Concepci\'on, 160-C Concepci\'on, Chile}
\affiliation{Center for Optics and Photonics, Universidad de Concepci\'on, Concepci\'on, Chile}
\affiliation{MSI-Nucleus for Advanced Optics, Universidad de Concepci\'on, Concepci\'on, Chile}

\author{Guilherme B. Xavier}
\affiliation{Center for Optics and Photonics, Universidad de Concepci\'on, Concepci\'on, Chile}
\affiliation{MSI-Nucleus for Advanced Optics, Universidad de Concepci\'on, Concepci\'on, Chile}
\affiliation{Departamento de Ingenier\'{\i}a El\'ectrica, Universidad de Concepci\'on, 160-C Concepci\'on, Chile}

\author{Gustavo Lima}
\affiliation{Departamento de F\'{\i}sica, Universidad de Concepci\'on, 160-C Concepci\'on, Chile}
\affiliation{Center for Optics and Photonics, Universidad de Concepci\'on, Concepci\'on, Chile}
\affiliation{MSI-Nucleus for Advanced Optics, Universidad de Concepci\'on, Concepci\'on, Chile}

\author{Ad\'{a}n~Cabello}
\affiliation{Departamento de F\'{\i}sica Aplicada II, Universidad de Sevilla, E-41012 Sevilla, Spain}


\begin{abstract}
We report a method that exploits a connection between quantum contextuality and graph theory to reveal any form of quantum contextuality in high-precision experiments. We use this technique to identify a graph which corresponds to an extreme form of quantum contextuality unnoticed before and test it using high-dimensional quantum states encoded in the linear transverse momentum of single photons. Our results open the door to the experimental exploration of quantum contextuality in all its forms, including those needed for quantum computation.
\end{abstract}




\maketitle


\section{Introduction}


Quantum theory (QT) is in conflict with the assumption that measurement outcomes correspond to preexisting properties that are not affected by compatible measurements \cite{Specker60,Bell66,KS67}. This conflict is behind the power of quantum computation \cite{HWVE14,DGBR15,RBDOB15} and quantum secure communication \cite{Ekert91,BHK05}, and can be experimentally tested through the violation of noncontextuality (NC) inequalities \cite{Cabello08}. These NC inequalities allow the observation of different forms of contextuality that cannot be revealed through Bell's inequalities: Contextuality with qutrits \cite{KCBS08}, quantum-state-independent contextuality \cite{Cabello08,BBCP09,YO12}, contextuality needed for universal fault-tolerant quantum computation with magic states \cite{HWVE14}, and absolute maximal contextuality \cite{ATC15} are just some examples. This variety of forms of contextuality leads to the question of how we can explore them theoretically and experimentally and, more precisely, to the following questions: (i) Is there a systematic way to explore all forms of quantum contextuality? (ii) What is the simplest way to experimentally test them?

Recently, there has been great progress towards solving both problems. On one hand, it has been shown that there is a one-to-one correspondence between graphs and quantum contextuality \cite{CSW14}: The figure of merit of any NC inequality can be converted into a positive combination of correlations $S$ to which one can ascribe a graph $G$, the so-called exclusivity graph of $S$. The maximum value of $S$ for noncontextual hidden variable theories (NCHVTs) is given by a characteristic number of $G$, the independence number $\alpha(G)$. The maximum in QT (or an upper bound to it) is given by another characteristic number of $G$, the Lov\'asz number $\vartheta(G)$, which has the advantage of being easy to compute. More interestingly in connection to question (i) is that, reciprocally, for any graph $G$ there is always a quantum experiment such that its maximum for NCHVTs is $\alpha(G)$ and its tight maximum in QT is $\vartheta(G)$ \cite{CSW14}. This provides a possible approach to solve problem (i), as it shows that all possible forms of quantum contextuality are encoded in graphs, so, by systematically studying these graphs, we can study all forms of quantum contextuality. In particular, by identifying graphs with specific properties, we can single out experiments with the corresponding quantum contextuality. Furthermore, other characteristic numbers of $G$ are associated with properties such as whether the quantum violation is state independent \cite{RH14,CKB15}.

Problem (ii) is also considered in Ref.\ \cite{CSW14}, where it is proven that orthogonal unit vectors can be assigned to adjacent vertices of $G$, satisfying that $\vartheta(G)=\sum_{i} |\langle u_i | \psi \rangle|^2$, for a particular unit vector $|\psi\rangle$. The vectors $\{u_i\}$ provide a so-called Lov\'asz optimum orthonormal representation of the complement of $G$ with handle $|\psi\rangle$ \cite{CDLP13}. This representation shows that the maximum quantum value of $S$ can be achieved by preparing the system in the quantum state $|\psi\rangle$ and projecting it on the different $|u_i\rangle\langle u_i|$.

However, a nontrivial problem remains, namely, how to carry out an experiment involving only compatible measurements and revealing the contextuality given by the graph. A solution to this problem has been recently presented in Ref.\ \cite{Cabello16}. There, it is shown that, for any $G$, there is always an experiment involving only compatible observables whose noncontextual and quantum limits are equal to the corresponding characteristic numbers of $G$. The proposed solution presents an additional advantage that is connected to problem (ii): It only requires testing two-point correlations. This suggests that it is possible to develop a new generation of contextuality tests, with a higher control of the experimental imperfections, and achieve conditions much closer to the ideal ones than those achieved in previous tests based on three-point correlations \cite{KZGKGCBR09,ARBC09}.

The aim of this work is to show that we can combine the theoretical results of Refs.\ \cite{CSW14,Cabello16}, with state-of-the-art experimental techniques for preparing and measuring high-dimensional photonic quantum systems \cite{Neves05, Lima09, Lima10, Lima11, Lima13, CEGSXLC14, CAEGCXL14, Dardo15, Heptagon15} into a method capable of systematically exploring all possible forms of quantum contextuality.


\section{Maximum quantum contextuality}


The method presented here is general. However, in order to show its power, we will focus on solving a particular problem: identifying and experimentally testing the simplest scenario in which the maximum quantum contextuality is larger than in any simpler previously studied scenario.

For identifying it, we consider a specific measure of contextuality introduced in Ref.\ \cite{ATC15}, which is specially useful when using graphs, namely, the ratio $\vartheta(G)/\alpha(G)$. Then, we study all graphs with a fixed number $n$ of vertices. For each $n$, we identify the graph with the largest $\vartheta(G)/\alpha(G)$. For that, we benefit from the exhaustive database developed in Ref.\ \cite{EDLPBA12}.

We observe that, for $n=5$ (the minimum $n$ for which quantum contextuality exists), the maximum of $\vartheta(G)/\alpha(G)$ is $\sqrt{5}/2 \approx 1.118$ and corresponds to a well-studied case, the maximum quantum violation of the Klyachko-Can-Binicio\u{g}lu-Shumovsky inequality (KCBS) \cite{KCBS08}, which is the simplest NC inequality violated by qutrits. For $n=6$ and $n=7$, the maximum of $\vartheta(G)/\alpha(G)$ is still $\sqrt{5}/2$. This shows that the maximum quantum contextuality for these values of $n$ is just a variant of the one in the KCBS inequality. For $n=8$, the maximum of $\vartheta(G)/\alpha(G)$ is $2 (2 - \sqrt{2}) \approx 1.172$ and also corresponds to a well-known case, the maximum quantum violation \cite{Tsirelson80} of the Clauser-Horne-Shimony-Holt (CHSH) Bell inequality \cite{CHSH69}, which has been recently reached in experiments \cite{PJCCK15}.

The fact that, by using this measure of contextuality and considering an increasing $n$, we have recovered the two most emblematic examples of quantum contextuality confirms the interest of the problem of identifying the graphs with maximum $\vartheta(G)/\alpha(G)$ for fixed $n$. As $n$ grows, the number of nonisomorphic graphs grows enormously and the exhaustive study of all of them becomes increasingly difficult. To our knowledge, such a comprehensive study of $\alpha(G)$ and $\vartheta(G)$ has been achieved only up to $n=12$ \cite{EDLPBA12}. Similar explorations suggest that this approach might be feasible up to $n=14$.

\begin{figure}[tb]
\includegraphics[trim = 5.8cm 8.8cm 5.8cm 8.8cm,clip,width=5.2cm]{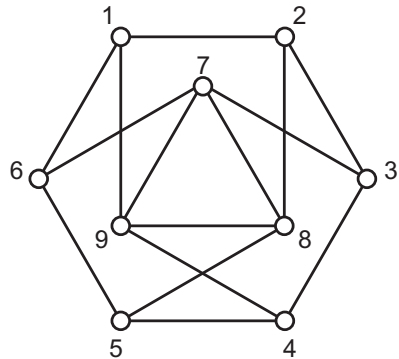}
\caption{``Fisher 9,'' $F_9$, the graph corresponding to the largest quantum contextuality, measured by $\vartheta(G)/\alpha(G)$, attainable with any graph with less than ten vertices. Vertices correspond to states and adjacent vertices to orthogonal states. \label{Fig1}}
\end{figure}

Interestingly, we have found that, for $n=9$, the maximum of $\vartheta(G)/\alpha(G)$ is already larger than the one in the CHSH inequality and that higher values of $n$ do not improve this maximum substantially \cite{EDLPBA12}. For $n=9$, this maximum is $11/9 \approx 1.222$ and only occurs for one graph, the graph in Fig.\ \ref{Fig1}. We will call this graph ``Fisher 9,'' $F_9$, since some of its properties were first pointed out in Ref.\ \cite{Fisher94}. This graph is also mentioned in Refs.\ \cite{Rubalcaba05,SU13}. However, to our knowledge, $F_9$ has not been mentioned in relation with QT.


To identify the minimum quantum dimension, the initial state, and the measurements needed to obtain the maximum quantum contextuality associated with $F_9$, we have to find a Lov\'asz-optimum orthonormal representation of the complement of $F_9$ with the smallest possible dimension. No such representation exists in Hilbert spaces of dimension smaller than four. We have found one which is particularly simple in dimension four and only contains states of the canonical basis and Hardy states \cite{Hardy93}. This representation is the following:
\begin{equation}
\begin{split}
\label{loor}
\langle \psi |= \frac{1}{\sqrt{3}}(1,1,1,0),\;\;\;\;\;\;\langle u_1 |= \frac{1}{\sqrt{3}}(1,1,0,1), \\
\langle u_2 |= \frac{1}{\sqrt{3}}(1,0,1,-1),\;\;\;\;\;\; \langle u_3 |= \frac{1}{\sqrt{3}}(0,1,1,1),\\
\langle u_4 |= \frac{1}{\sqrt{3}}(1,1,0,-1),\;\;\;\;\;\;\langle u_5 |= \frac{1}{\sqrt{3}}(1,0,1,1),\\
\langle u_6 |= \frac{1}{\sqrt{3}}(0,1,1,-1),\;\;\;\;\;\;\langle u_7 |= (1,0,0,0),\\
\langle u_8 |= (0,1,0,0),\;\;\;\;\;\;\langle u_9 |= (0,0,1,0).
\end{split}
\end{equation}

The next step is to identify an experimentally testable NC inequality containing only correlations among compatible measurements and such that its noncontextual bound is $\alpha(F_9)=3$ and its maximum quantum violation is $\vartheta(F_9)=11/3$, and is achievable with the state $|\psi\rangle$ and measuring the projectors $i=|u_i\rangle \langle u_i|$. For that, we use the result in Ref.\ \cite{Cabello16}, according to which one inequality with those properties is the following equation:
\begin{equation}
\label{main}
{\cal S} \equiv \sum_{i \in V(F_9)} P(1|i) - \sum_{(i,j) \in E(F_9)} P(1,1|i,j) \stackrel{\mbox{\tiny{NCHVTs}}}{\leq} 3,
\end{equation}
where $V(F_9)$ is the vertex set of $F_9$, $E(F_9)$ is the edge set of $F_9$, and $P(1,1|i,j)$ is the joint probability of obtaining outcomes $1$ and $1$ when we measure $i$ and $j$. One can check that, if we prepare the quantum state $|\psi\rangle$ and measure $i=|u_i\rangle \langle u_i|$ and $j=|u_j\rangle \langle u_j|$, then ${\cal S}=\vartheta(F_9)=11/3$.


\begin{figure}[tb]
 \includegraphics[width=0.40 \textwidth]{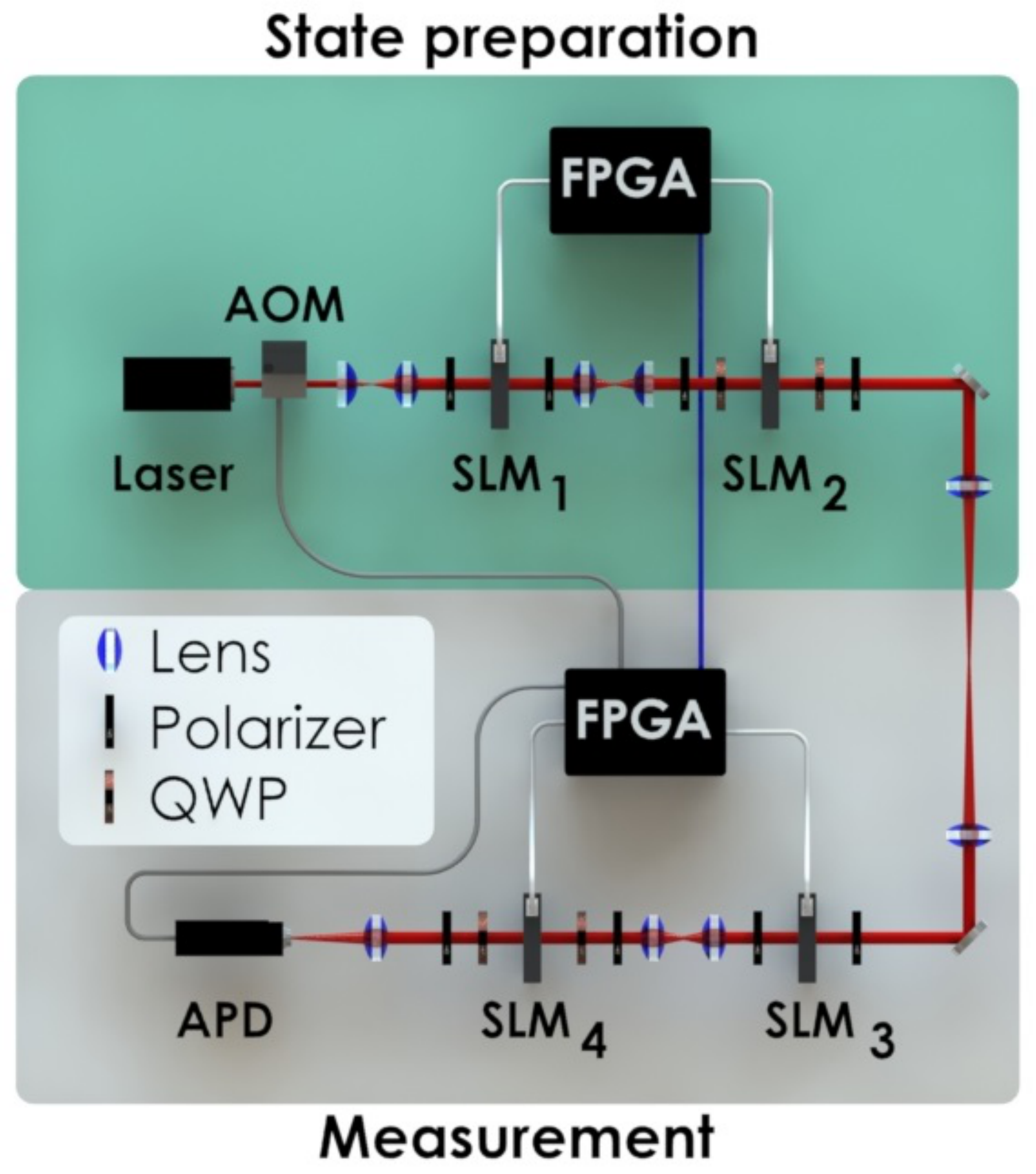}
 \caption{Experimental setup for preparing and measuring spatial qudits encoded on the linear transverse momentum of weak coherent states. At the state preparation stage, the encoding of each state is implemented by two spatial light modulators (SLMs). The measurement is implemented using two SLMs but now combined with a pointlike single-photon detector (APD). The experimental protocol is automatically controlled by the two field programmable gate array (FPGA) electronic modules. They work synchronously, as indicated by the blue cable connecting them. See main text for details. \label{Fig2}}
\end{figure}


\section{Experimental setup}


In order to test inequality (\ref{main}), we use the linear transverse momentum of single photons. This approach has been successfully used to produce and manipulate high-dimensional photonic quantum systems \cite{Neves05,Lima09, Lima10, Lima11, Lima13, CEGSXLC14, CAEGCXL14, Dardo15, Heptagon15}. The setup used in our experiment is depicted in Fig.\ \ref{Fig2} and exploits the idea in Ref.\ \cite{Cabello16} for testing two-point correlations using quantum systems. To obtain two-point correlation probabilities needed to test inequality (\ref{main}), we first perform a measurement of $i=|u_i\rangle \langle u_i|$ on a system prepared in state $|\psi\rangle$. If the result is $1$, we then prepare a new system in the state $|u_i\rangle$ and perform the measurement of $j=|u_j\rangle \langle u_j|$. If the initial result of $|u_i\rangle \langle u_i|$ is $0$, we do not need to perform, in principle, any further measurement since we just need $P(1|i)$ and $P(1,1|i,j)$ to test inequality (\ref{main}).

However, the assumption of noncontextuality leading to inequality (\ref{main}) is legitimate insofar as the statistics of the measurement outcomes are not perturbed by previous measurements, and it is then important to test that this condition is achieved in our experiment. This test requires that one also measures $P(0,1|i,j)$. Thus, if the initial result of $|u_i\rangle \langle u_i|$ measurement is $0$, we also prepare $|u_i^\bot\rangle$ (defined next), which is the state obtained after a projective measurement of $i$ with outcome $0$ on a system initially prepared in state $|\psi\rangle$. We then measure $j$.

Consequently, our experimental setup consists of two parts: the state preparation (SP) stage and the measurement stage. At the SP stage the single-photon regime is achieved by heavily attenuating optical pulses, which are generated with an acousto-optical modulator (AOM) placed at the output of a continuous-wave laser operating at 690 nm. Well-calibrated attenuators are used to set the average number of photons per pulse to $\mu=0.16$. In this case, the probability of having non-null pulses, i.e., of having pulses containing at least one photon, is $P(\mu=0.16|n\geq1)=14.8\%$. Pulses containing only one photon are the vast majority of the non-null pulses generated and account for 92.2$\%$ of the experimental runs. The probability of multiphoton events is negligible as it is $\simeq$ 1.2$\%$. Therefore, our source can be seen as a good approximation to a nondeterministic single-photon source, which is commonly adopted in quantum key distribution \cite{Gisin}.

In order to prepare the $d$-dimensional quantum states we employ the linear transverse momentum of single photons. The generated photons are sent through diffractive apertures addressed in spatial light modulators (SLMs), and the four-dimensional state required in the experiment is defined by addressing four parallel slits in the SLMs for the photon transmission. All slits in each modulator have the same physical dimension, that is, each has a width of 96 $\mu$m and an equal center-to-center separation. In this case, the state of the transmitted photons is given by

\begin{equation}
\label{state}
|\psi_{\mathrm{expt.}}\rangle= \frac{1}{\sqrt{C}}\sum_{l=-\frac{3}{2}}^{l=\frac{3}{2}}\sqrt{t_{l}}e^{i\phi_{l}}|l\rangle,
\end{equation}
where $|l\rangle$ represents the state of a photon transmitted by the $l$th slit. $t_l$ ($\phi_l$) is the transmissivity (phase) defined for each slit and $C$ the normalization constant \cite{Neves05,Lima09}.

Two SLMs are used at each stage. In the SP stage the first SLM controls the real part of the coefficients of the generated states, while the second SLM their phases \cite{CAEGCXL14}. Sets of lenses are employed to ensure that each SLM is placed on the image plane of the next one. In the measurement stage the state projection is performed using a second pair of SLMs and a pointlike avalanche photo-detector (APD). After the last modulator, the attenuated laser beam is focused at the detection plane. The pointlike detector is constructed with a small circular pinhole (10 $\mu$m diameter), followed by a silicon single-photon avalanche photodetector (APD), which is then positioned at the center of the interference pattern. In this configuration the detection probability is proportional to $|\langle{\psi_{\mathrm{expt.}}}|k\rangle|^2$, where $|k\rangle$ is the state at the measurement stage (see Ref.\ \cite{Lima13} for details).


\section{Measurement results}


\begin{figure}[tb]
\includegraphics[trim = 3.9cm 7.6cm 3.5cm 7.6cm,clip,width=6.7cm]{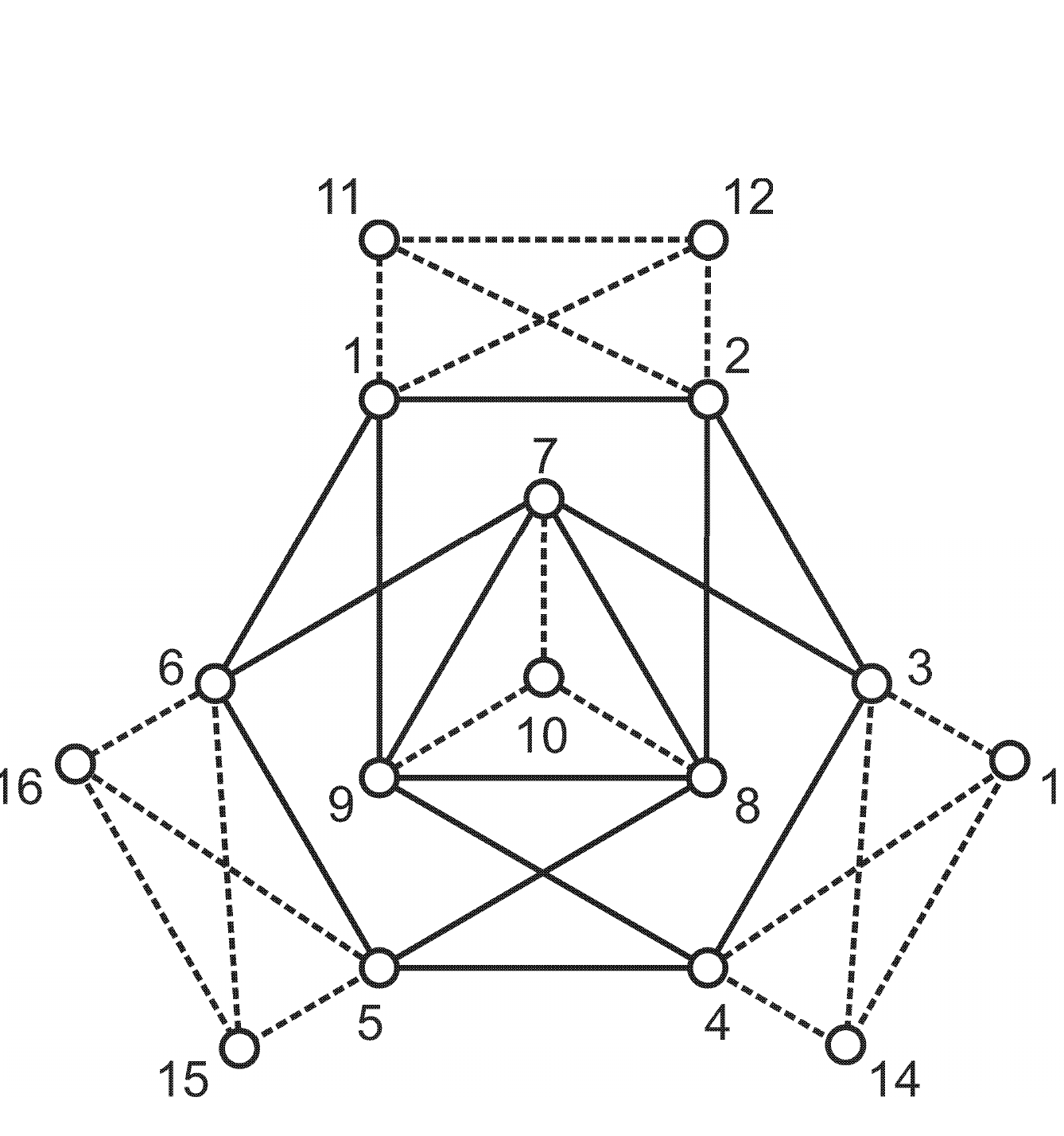}
\caption{Graph extending the graph in Fig.\ref{Fig1} with seven new vertices (vertices 10--16) such that each of the vertices of Fig.\ref{Fig1} with vertices 1--9 now belongs to a clique of size four (which represents an orthogonal basis of the four-dimensional Hilbert space). Adjacent vertices correspond to orthogonal states. Dashed lines indicate new orthogonalities with respect to Fig.\ref{Fig1}. \label{Fig}}
\end{figure}

To properly determine the experimental probabilities required to test inequality (\ref{main}) we have extended $F_9$ into a larger graph in which every vertex belongs to exactly one clique of size four (i.e., an orthogonal basis). The extended graph is shown in Fig.\ \ref{Fig}, and the new vertices correspond to the following states:
\begin{equation}
\begin{split}
&\langle u_{10} |= (0,0,0,1), \\
&\langle u_{11} |= \frac{1}{\sqrt{3}}(0,-1,1,1),\;\;\;\;\;\;\langle u_{12} |= \frac{1}{\sqrt{3}}(-1,1,1,0), \\
&\langle u_{13} |= \frac{1}{\sqrt{3}}(1,0,-1,1),\;\;\;\;\;\;\langle u_{14} |= \frac{1}{\sqrt{3}}(1,-1,1,0), \\
&\langle u_{15} |= \frac{1}{\sqrt{3}}(1,1,-1,0),\;\;\;\;\;\;\langle u_{16} |= \frac{1}{\sqrt{3}}(-1,1,0,1).
\end{split}
\end{equation} Notice that these states allow us to measure each observable $i$ using always the same orthogonal basis $\{i,i',i'',i'''\}$ independently of the context.  For instance, the probabilities $P(1|i)$ and $P(0|i)$ are calculated from the experimental data as follows:
\begin{equation}
\begin{split}
P(1|i) = & \frac{N(i)}{N(i)+N(i')+N(i'')+N(i''')},\\
P(0|i) = & \frac{N(i')+N(i'')+N(i''')}{N(i)+N(i')+N(i'')+N(i''')},
\end{split}
\end{equation}
where $N(i)$ is the number of counts corresponding to outcome $i$ and $\{i,i',i'',i'''\}$ is an orthogonal basis.

As we mentioned above, in the SP stage we prepare the state $\langle\psi_{\mathrm{expt.}}|$ and then project it on $i=|u_i\rangle \langle u_i|$ at the measurement stage. If the outcome is 1, we prepare the state $\langle u_i|$ and now the measured projector is $j=|u_j\rangle \langle u_j|$. On the other hand, if the output is 0, the state $\langle u_i^\bot|$ is prepared and projected on $j=|u_j\rangle \langle u_j|$. The states $\langle u_i^\bot|$ are defined by

\begin{equation}
 |u_i^\bot\rangle= \frac{(\openone-|u_i\rangle \langle u_i|) |\psi\rangle}{\left[\langle \psi | (\openone-|u_i\rangle \langle u_i|) |\psi\rangle \right]^{\frac{1}{2}}},
\end{equation}
where $\openone$ denotes the four-dimensional identity matrix. More specifically, in our experiment these states are given by
\begin{equation}
\begin{split}
&\langle u_1^\bot |= \frac{1}{\sqrt{15}}(1,1,3,-2),\;\;\;\;\;\;\langle u_2^\bot |= \frac{1}{\sqrt{15}}(1,3,1,2), \\
&\langle u_3^\bot |= \frac{1}{\sqrt{15}}(3,1,1,-2),\;\;\;\;\;\;\langle u_4^\bot |= \frac{1}{\sqrt{15}}(1,1,3,2), \\
&\langle u_5^\bot |= \frac{1}{\sqrt{15}}(1,3,1,-2),\;\;\;\;\;\;\langle u_6^\bot |= \frac{1}{\sqrt{15}}(3,1,1,2), \\
&\langle u_7^\bot |= \frac{1}{\sqrt{2}}(0,1,1,0),\;\;\;\;\;\;\;\;\;\;\langle u_8^\bot |= \frac{1}{\sqrt{2}}(1,0,1,0), \\
&\langle u_9^\bot |=\frac{1}{\sqrt{2}} (1,1,0,0).
\end{split}
\end{equation}

During the measurement procedure, the SP and measurement stages run in an automated fashion, controlled and synchronized by the two FPGA electronic modules at a rate of 30 Hz. One module is placed at the SP stage while the other one controls the measurement stage, reads the APD output and sends the results to a personal computer for further processing.


\begin{figure}[tb]
 \includegraphics[width=0.37 \textwidth]{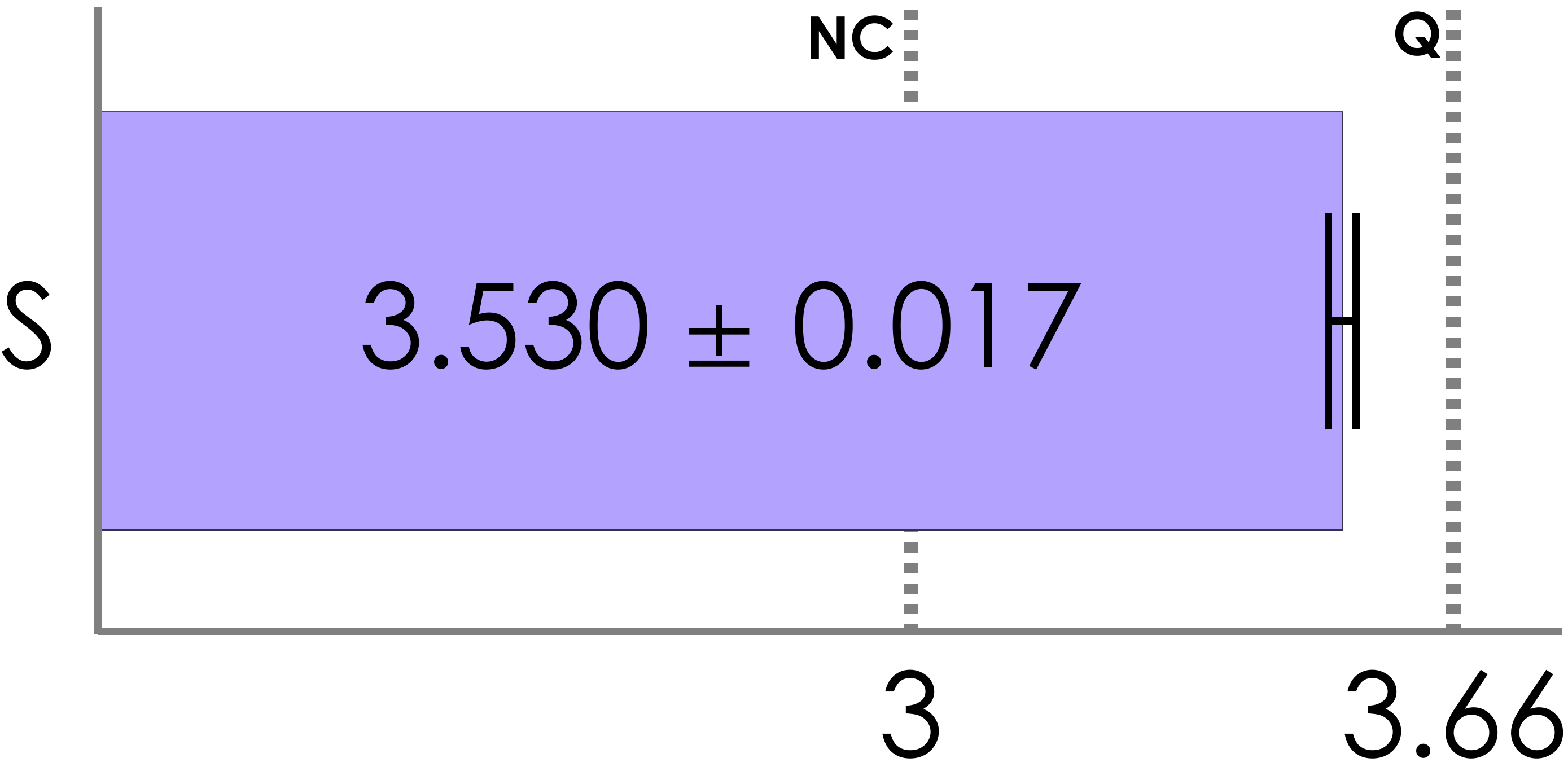}
 \caption{Experimentally measured violation of NC inequality (\ref{main}). NC indicates the limit for NCHV theories and Q the maximum allowed by QT.} \label{Fig3}
\end{figure}


From the recorded data, we extract the probabilities to test the violation of inequality (\ref{main}) and also to verify that there is no signaling between the first measurement associated to $|u_i\rangle \langle u_i|$ and the second measurement associated to $|u_j\rangle \langle u_j|$. The experimental value of ${\cal S}$ is depicted in Fig.\ \ref{Fig3} and shows a violation of inequality (\ref{main}) by over 30 standard deviations. The maximum quantum value of ${\cal S}$ is not reached due to intrinsic experimental misalignments and the detector's dark counts that produce, e.g., nonzero values for the probabilities $P(1,1|i,j)$. Nevertheless, notice that our experimental value corresponds to a degree of contextuality that surpasses the maximum attainable through the quantum violation of the KCBS or CHSH inequalities.


\begin{figure*}[t]
 \includegraphics[width=1 \textwidth]{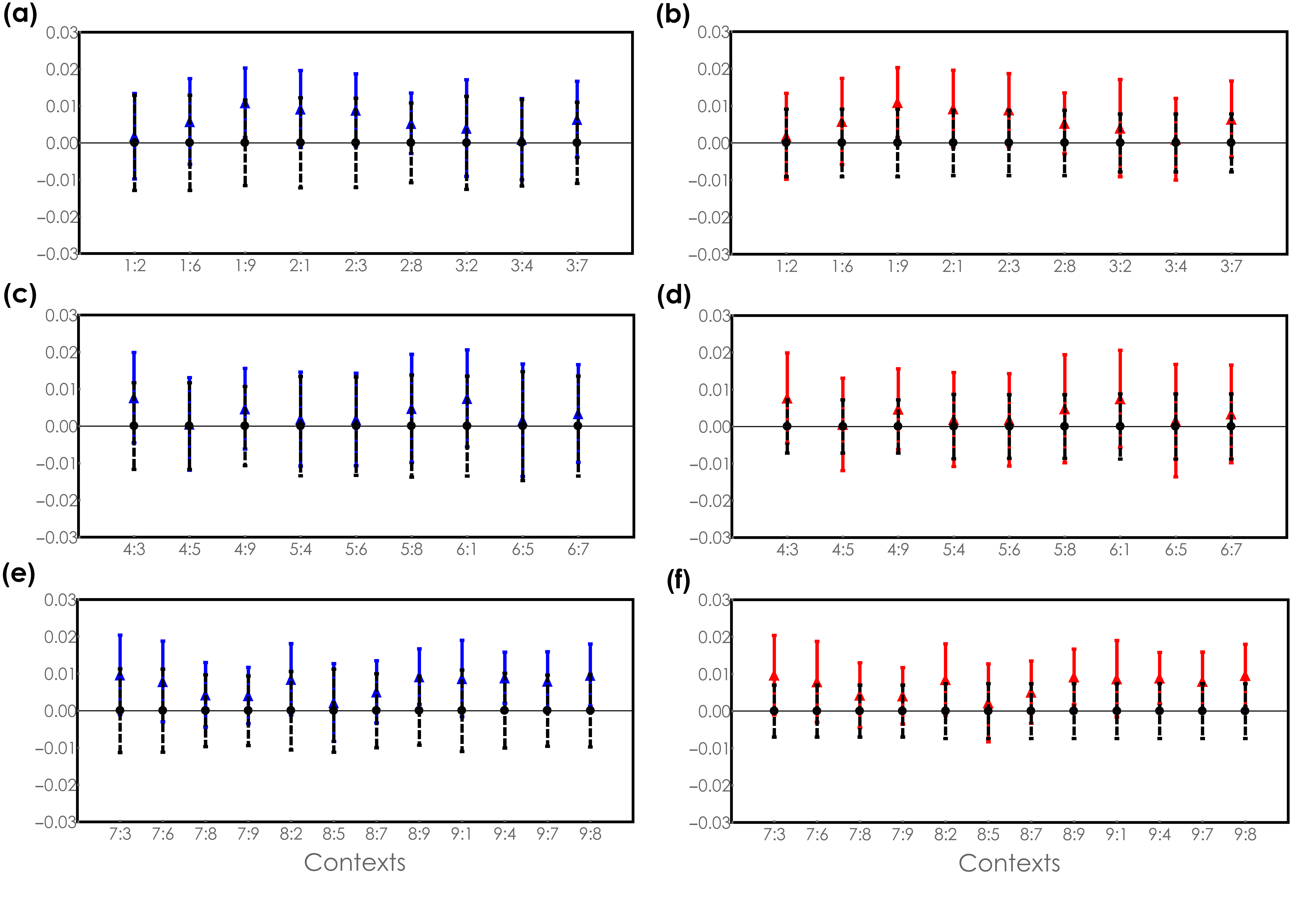}
 \caption{(a), (c), and (e) Experimental values for $\epsilon_{\_,0|i,j}$ (in blue) and $\epsilon_{0,\_|i,j}$ (in black), defined in Eqs.\ (\ref{Eq9}) and (\ref{Eq10}). The notation is $i:j$. (b), (d), and (f) Experimental values for $\epsilon_{\_,1|i,j}$ (in red) and $\epsilon_{1,\_|i,j}$ (in black). As expected, the range of the error bars in blue and red is within the one of the error bars for the expected zeros (in black). The average value of both $\epsilon_{\_,0|i,j}$ and $\epsilon_{\_,1|i,j}$ is $0.006 \pm 0.011$. \label{Fig4}}
\end{figure*}


In order to test that there is no signaling between measurements $i$ and $j$, we have used
\begin{equation} \label{Eq9}
\begin{split}
\epsilon_{\_,0|i,j} \equiv & |P(0|j) - P(0,0|i,j)- P(1,0|i,j)|, \\
\epsilon_{\_,1|i,j} \equiv & |P(1|j) - P(0,1|i,j)- P(1,1|i,j)|
\end{split}
\end{equation}
to measure how the first measurement, $i$, affects the statistics of the second measurement, $j$.
Our purpose is to certify that, for all $i$ and $j$ in inequality (\ref{main}), the experimental values of $\epsilon_{\_,0|i,j}$ and $\epsilon_{\_,1|i,j}$ are compatible with zero and have the same error as the experimental quantities
\begin{equation} \label{Eq10}
\begin{split}
\epsilon_{0,\_|i,j} \equiv & |P(0|i) - P(0,0|i,j)- P(0,1|i,j)|, \\
\epsilon_{1,\_|i,j} \equiv & |P(1|i) - P(1,0|i,j)- P(1,1|i,j)|,
\end{split}
\end{equation}
which, according to causality, are zero, but whose error gives the experimental precision with which we can determine a zero within our experiment. The idea of this approach is to show that in our work there is the same signaling between past and future measurements than between future and past measurements. If we assume that the latter is zero, from the obtained results we can conclude that the experimental data are compatible with the assumption that the former is zero. To obtain $\epsilon_{\_,0|i,j},
\epsilon_{\_,1|i,j}, \epsilon_{0,\_|i,j}$, and $\epsilon_{1,\_|i,j}$ we use that
\begin{equation} \label{Eq6}
\begin{split}
 P(1,1|i,j) = & P(1|i)P(1|j), \\
 P(0,1|i,j) = & P(0|i)P(1|j), \\
 P(1,0|i,j) = & P(1|i)- P(1,1|i,j),\\
 P(0,0|i,j) = & P(0|i)- P(0,1|i,j).
\end{split}
\end{equation}

The experimental values for $\epsilon_{\_,0|i,j}$, $\epsilon_{\_,1|i,j}$, $\epsilon_{0,\_|i,j}$, and $\epsilon_{1,\_|i,j}$ for all pairs $(i,j)$ of measurements used to test inequality (\ref{main}) are shown in Fig.\ \ref{Fig4}. They show that in our experiment the influences of the first measurements on the second ones are negligible.


\section{Conclusions}


Being so fundamental for quantum theory, quantum computation, and quantum secure communication, it is surprising how little effort has been made to experimentally investigate quantum contextuality beyond Bell's inequalities. Here we have demonstrated a tool for exploring, theoretically and experimentally, quantum contextuality in all its forms. We have described all the steps of a method to, first, identify interesting forms of contextuality and, then, to design and perform precise experiments to reveal them. Our approach is universal and can be applied to study any form of quantum contextuality. In particular, it opens the possibility of experimentally testing the contextuality needed for quantum computation \cite{HWVE14,DGBR15,RBDOB15}. In addition, we have shown that the approach is useful in itself, since it is capable of revealing interesting cases unnoticed before. Its only limitations are our ability to explore large graphs or perform experiments requiring a large number of two-point correlations. Moreover, we have seen that this approach leads to photonic tests allowing a better control of the imperfections and higher-quality results (closer to the predictions of quantum theory under ideal conditions) than previous experiments. In particular, we have verified that it allows for experiments in which the signaling between past and future measurements is negligible. In summary, although there is still work to be done for closing loopholes \cite{G10} and improving the analysis of the experimental data \cite{Winter14,DKL15}, our results indicate that we already have powerful tools for exploring a fundamental part of quantum theory.


\begin{acknowledgments}
We thank A.\ J.\ L\'opez-Tarrida and J.\ R.\ Portillo for discussions. This work was supported by FONDECYT Grants No.\ 1160400, No.\ 11150324, No.\ 11150325, and No.\ 1150101, Milenio Grant No.\ RC130001, PIA-CONICYT Grant No.\ PFB0824, the FQXi large grant project ``The Nature of Information in Sequential Quantum Measurements,'' Project No.\ FIS2014-60843-P ``Advanced Quantum Information'' (MINECO, Spain) with FEDER funds, and the project ``Photonic Quantum Information'' (Knut and Alice Wallenberg Foundation, Sweden). J.C.\ and J.F.B.\ acknowledge the support of CONICYT. A.C.\ thanks the CEFOP for its hospitality.
\end{acknowledgments}


\end{document}